\documentstyle[12pt,aps,epsfig]{revtex}
\newcommand{\snn}{\sqrt{s_{\scriptscriptstyle nn}}}
\begin{document}
\draft
\tighten
\preprint{}
\title{Systematic Study of Elliptic Flow at RHIC Energy
}
\author{P. K. Sahu$^{1}$, N. Otuka$^{2}$ and A. Ohnishi$^{2}$ 
}
\address{$^{1}$Institute of Physics, Bhubaneswar 751005, India
\\
$^{2}$Division of Physics, Graduate School of Science, Hokkaido University\\
 Sapporo 060-0810, Japan\\
}
\maketitle

\abstract{
We study the elliptic flow systematically from SIS to RHIC energies 
in a realistic dynamical cascade model.
We compile our results with the recent data from STAR and PHOBOS 
experiments on elliptic flow of charged particles in Au + Au 
collisions at RHIC energy.
From the analysis of elliptic flow as a function of different
dynamical variables such as transverse momenta, pseudorapidity and
centrality at RHIC energy, we found a good
fitting with data at 1.5 times a scaling factor to our simulation
model, which characterizes that the model is required to have extra 
pressure generated from the subsequent parton scatterings.
In energy dependence of elliptic flow,
we observe a re-hardening nature at RHIC energies,
which may probably signal
the possible formation of quark-gluon plasma.
}
\vskip 0.2 in

\section{Introduction}
%%%%%%%%%%%%%%%%%%%%%%%%%%
% General Introduction
%%%%%%%%%%%%%%%%%%%%%%%%%%
The prime aim of ultra-relativistic heavy-ion collisions is
to understand the nature of quantum chromo dynamics 
under extreme conditions of density and temperature. 
In such extreme conditions, it is expected that nuclear matter undergo
a phase transition to quark-gluon plasma (QGP).
At present it is of great interest to study the nature of this plasma 
and understand the phase transition between 
hadron and QGP phases.
For that reason,
various high energy 
heavy-ion collision experiments have been carried out
at SIS, AGS, SPS and RHIC energies, and will start at LHC around 2007.
Very recently, the new qualitative data have been reported 
from RHIC energy experiments at BNL.
These are glimpse of the wealth of physics to be extracted from
four experiments BRAHMS, PHENIX, PHOBOS and STAR at RHIC accelerator.

%%%%%%%%%%%%%%%%%%%%%%%%%%
% Flow
%%%%%%%%%%%%%%%%%%%%%%%%%%
At RHIC energy,
soon after the collisions of heavy nuclei,
a huge numebr of particles are produced and move collectively.
The collective motion and the behavior of these particles
are called as {\em flow}.
Recently these flow data have been reported at AGS and SPS
besides RHIC energy.
At AGS energies, the sideward and elliptic flow are well described
by dynamical microscopic simulation models~\cite{sahu00}
in non-central Au+Au collisions.
Also, the elliptic flow in non-central and strong radial flow in central 
Pb+Pb collisions are observed at the SPS~\cite{na49}.
For non-central collisions, the initial nucleus-nucleus overlap has an
almond or elliptic shape.
This initial almond shaped overlap region expands and becomes more spherical,
quenching the driving force that produces the elliptic flow.
The pressure gradient and its anisotropy are much larger in the initial stage,
hence the elliptic flow give more precise information
of the initial thermalization and equation of state.
At RHIC where deconfined phase is expected to emerge,
the elliptic flow would be more sensitive
to the parton re-scatterings and thermalization degree 
in the initial stage 
than to the later hadronic equation of state.
So, the information about the formation of QGP can be drawn from the 
measure of the final flows, e.g. radial and elliptic flow of the 
produced particles.

%%%%%%%%%%%%%%%%%%%%%%%%%%
% Flow
%%%%%%%%%%%%%%%%%%%%%%%%%%
For radial flows, there is a strong evidence
that the hadron transverse mass spectra get much stiffer
than SPS energy~\cite{NuXu}.
The stiffness can be realized 
as follows.
According to the large level density of hadrons, the hadronic matter
is expected to be softer.
However, this phenomena can not persist 
at such high energies,
since hadrons are dissolved into quark and gluons in vacuum.
Therefore in that level, pressure grows rapidly and becomes stiffer as a
function of energy density.
In other words, we say re-hardening of transverse mass spectra
at RHIC energies~\cite{NuXu,otuka01} is due the probable formation of QGP.
This point may be quite premature, 
because the radial flow is not a direct observable but a quantity
extracted through theoretical model analayses,
and there may be some other mechanism such as $p_t$ broadning.

%%%%%%%%%%%%%%%%%%%%
% v2
%%%%%%%%%%%%%%%%%%%%
For non-central collisions,
the overlap geometry between two nuclei is lens or almond shaped.
As the initial lens-shape expands, it produces the elliptic flow.
The elliptic flow is the anisotropic emission of particles in- and out-of
reaction plane defined by the beam and the impact parameter directions.
Thus the momentum anisotropy can be translated from the spatial 
anisotropy in the presence of strong re-scattering
and elliptic flow is sensitive to number of interaction.
One can measure this by measuring the second Fourier coefficient in the 
azimuthal distribution of particles with respect to reaction plane and is 
usually characterized by the particle momenta distribution~\cite{star01},
\begin{equation}
v_2=<({p_x}^2-{p_y}^2)/(({p_x}^2+{p_y}^2)>.
\end{equation}
Also, the elliptic flow is influenced by the formation of QGP
in non-central collisions with function of beam energies,
since it depends on the early stages of the system evolution.
Then the question arises if the QGP is formed,
does it live longer at SPS or at RHIC ?
Recently, it has been estimated experimentally~\cite{alber95}
that the hard QGP phase is expected to live longer at RHIC than at the SPS.
If this is true, then the elliptic flow of the produced
particles should indicate this difference at the end.
Therefore, it is urgently required to estimate the incident energy 
dependence of the elliptic flow
more systematically in order
to derive the new physics.
Experimentally, it has been found that at higher energies, e.g., at AGS and
above, the coefficient $v_2>0$, the "in-plan" flow.
This fact has been verified and well described by the dynamical transport
model with mean field up to AGS energies~\cite{sahu00}.
In any case, whether the transport model has mean field or not, the elliptic
flow $v_2$ is positive at higher energies.
Recently,
the elliptic flow has been predicted to 
increase with beam energies
by RQMD~\cite{sorge95} as well as hydrodynamic models~\cite{kolb01}.

%%%%%%%%%%%%%%%%%%%%
% Purpose
%%%%%%%%%%%%%%%%%%%%
In this paper, 
we concentrate on the systematic study of elliptic flow,
because of two reasons,
(i) we have lots of quality data on elliptic flow from RHIC experiments,
and (ii) it is more fundamental to understand observables
which are sensitive to the scatterings among produced particles
in the initial stage.
Therefore, we make an analysis from SIS to RHIC energies,
and a detail discussion at RHIC energy
as functions of centrality, pseudorapidity and transverse momentum.

\section{Model}
%%%%%%%%%%%%%%%%%%%%%%%%%%
% JAM
%%%%%%%%%%%%%%%%%%%%%%%%%%
In this work, we analyze the elliptic flow systematically 
from SIS to RHIC energies using a dynamical hadron-string 
cascade simulation model, JAM~\cite{nara00}.
In this model, the initial primary collisions produce mini-jet partons
by using the eikonal approximation as in the HIJING model~\cite{wang94},
which later enter into string configurations. 
Then strings fragment to hadrons using the LUND fragmentation model 
constructed in the PYTHIA~\cite{pythia94} routine.
In JAM, partonic interactions between different mini-jets are not included.
However, at present very few models (no dynamical models) exist in the
literature which hold such a complicated treatments as 
partonic, string, and hadron multiple interactions at RHIC energy.
Thus it is worthwhile to consider the present model, JAM,
for the systematics of elliptic flow.

\section{Results}
\begin{figure}%[htbp]

\centerline{~{\psfig{figure=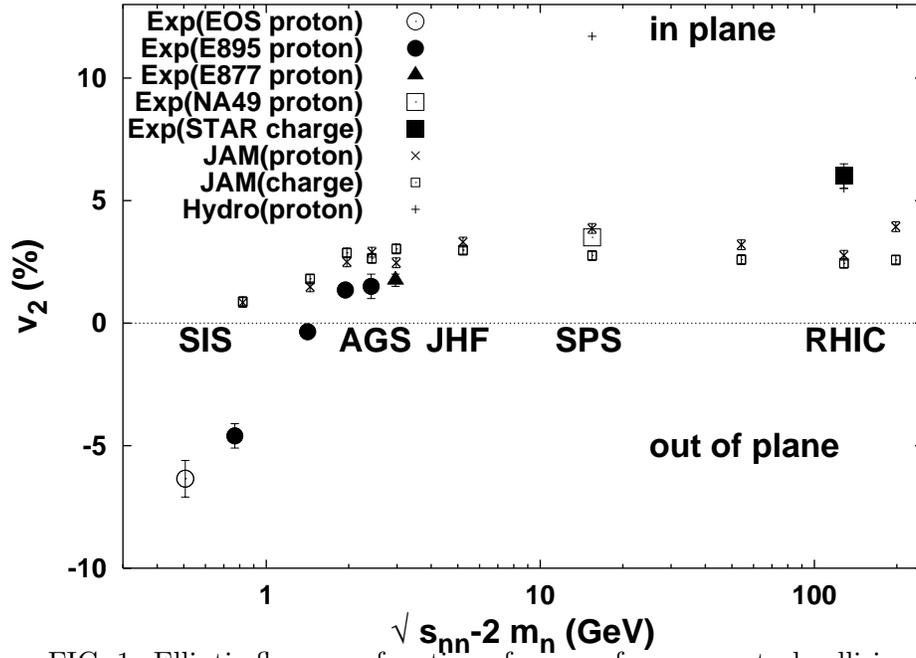,width=9cm,angle=-90}}~}
\caption{Elliptic flow as a function of energy for non-central collisions}
\label{Fig:Einc}
\end{figure}

%%%%%%%%%%%%%%%%%%%%%%%%%%%%%%%%%%%%%%%%%%%%
% Fig:Einc:  Incident Energy Dependence
%%%%%%%%%%%%%%%%%%%%%%%%%%%%%%%%%%%%%%%%%%%%
Figure~\ref{Fig:Einc} displays the calclulated results of $v_2$
at mid-rapidity
as functions of beam energy in JAM in comparison with experimental data.
For completeness, the result of hydrodynamic models\cite{kolb99} are 
displayed in the figure.
It is evident from the figure that the hydrodynamic model for protons
is very well on top of the STAR data~\cite{star02} at RHIC energy, 
where JAM under-predicts by a factor of 1/2 of data.
Whereas the situation is reverse at SPS, where JAM gives a reasonable 
description of data and hydrodynamic model fails to describe the data,
which overestimates the data by a factor of more than 2.
Since hydrodynamic models assume complete local thermalization
and QGP formation in the initial condition,
the agreement of hydrodynamics results with data may be suggesting
that the thermalized QGP is already formed in RHIC energy at mid-rapidities.
In JAM,
although it fails to explain mid-rapidity STAR data at RHIC energy,
it gives reasonable values up to SPS energies,
where hadronic matter is expected to dominate.
This shows that JAM lacks partonic interactions between mini-jets
which play essential roles in early thermalization, 
while hadron and string interactions are well implemented in this model.

%%%%%%%%%%%%%%%%%%%%%%%%
% Fig:Pt: Pt dependence
%%%%%%%%%%%%%%%%%%%%%%%%
In Fig.~\ref{Fig:Pt},
we display JAM results on elliptic flow of
charged particles as functions of transverse momenta
for minimum bias events at RHIC energy.
In this figure,
we notice that our model gives a good qualitative description of data.
However, the overall magnitudes are underprediced by a factor
of 1.2-1.5 for pions and charged particles.
If we multiply by a factor of 1.5 times the charged particle results, 
represented as thin dashed line in Fig.~\ref{Fig:Pt},
the data and model results are in an excellent agreement
till $p_t\sim$ 2 GeV/c.

Another interesting point to be noticed here is
that the calculated elliptic flow is sensitive to the particle masses
as a function of $p_t$.
The particles which are having smaller masses have higher values of elliptic
flow at small $p_t$,
such as pions and kaons and these are linear functions of $p_t$.
For higher particles masses such as protons, $v_2$ behaves non-linearly
with $p_t$.
Similar characteristics are observed in the STAR data~\cite{star01a} as well.

In contrast, hydrodynamic models show excellent agreement upto 
$p_t\sim 1.5$ GeV and in central and semi-central 
collisions~\cite{kolb01}, which is not shown in this figure. 
It fails at high $p_t$, due to saturation and onset of hard processes
and fails at peripheral collisions due to incomplete 
early-time thermalization.

%%%%%%%%%%%%%%%%%%%%%%%%%%%%%%%%%%%%
% Fig:Cent: centrality dependence
%%%%%%%%%%%%%%%%%%%%%%%%%%%%%%%%%%%%
Figure~\ref{Fig:Cent} shows
the centrality ($N_{ch}/N_{max}$) dependence of the elliptic flow.
The data are from STAR~\cite{star01} and PHOBOS~\cite{phobos02} experiments.
At central region,
our model fits much better (solid line) to data and fall 
off away at peripheral region.
Even at peripheral region, the experimental data are having large error bars,
especially those from PHOBOS experiment.
Again consistently, if we multiply a factor of 1.5 to our calculation
(dashed line), we could describe data well
in a wide range of impact parameters within the error bars. 
In comparison to hydrodynamic model at peripheral region, the hydrodynamic 
prediction overestimates the elliptic flow data in the most peripheral
region, which is not shown in the figure.
In over all, a large degree of thermalization are favored
in central collisions (in the early stages)
and fails at peripheral collisions.

\begin{figure}%[htbp]
\centerline{~{\psfig{figure=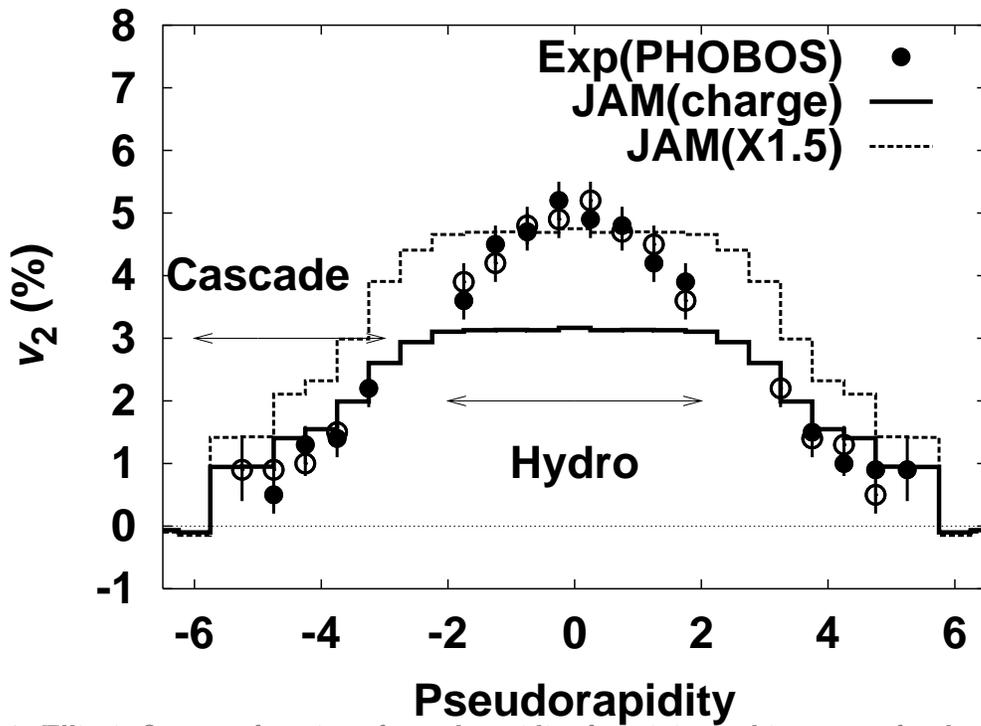,width=10cm,angle=-90}}~}
\caption{Elliptic flow as a function of pseudorapidity 
for minimum bias events for charged particles}
\label{Fig:Eta}
\end{figure}

%%%%%%%%%%%%%%%%%%%%%%%%
% Fig:Eta: eta dependence
%%%%%%%%%%%%%%%%%%%%%%%%
Finally, Fig.~\ref{Fig:Eta} displays
the calculated minimum bias elliptic flow of charged particles
as a function of pseudorapidity
in comparison with data from PHOBOS collaboration~\cite{phobos02}
at RHIC energy ($\snn = 130$ GeV).
In this figure we observe that
the calculated results agree with the data very well 
in the fragmentation region $(|\eta| > 2)$
labeled as "Cascade" in the figure.
If we multiply a factor of 1.5 to our results (dashed line)
as in previous figures,
we can describe the PHOBOS data at mid-rapidities.
On the other hand, a full 3D hydrodynamical model explains
the strong elliptic flow at mid-rapidities,
while it gives very flat ellptic flows
and consequently overestimates the data at large rapidities
for all reasonable initial conditions~\cite{hirano01}.
These findings implies
that well thermalized matter is produced at mid-rapidities,
where hydrodynamical evolution from QGP initial condition would be justified,
and that hadron-string gas still dominates at large pseudorapidity region,
where a jet-implimented hadron-string cascade model works well.

\section{Summary and Discussion}
%%%%%%%%%%%%%%
% Conclusion
%%%%%%%%%%%%%%
In this work, we have made systematic analyses of the elliptic flow
at beam energies ranging from SIS to RHIC,
and we have also discussed its dependence on 
transverse momentum, pseudorapidity and centrality at RHIC energy.

From the systematic analysis of elliptic flow at mid-rapidities 
with beam energy, which is more fundamental about the collision dynamics
with dynamical simulation models,
we have learned that the elliptic flow shows re-hardening behavior at RHIC
energy in non-central collisions.
The similar feature was observed from
the analysis of radial flow in the central collisions~\cite{NuXu,otuka01}.
A jet-implemented hadron-string cascade model, JAM,
reasonably describes the elliptic flows up to SPS energies, 
but it underestimates at RHIC
by a factor around two at the centrality where $v_2$ becomes maximum.
Transeverse momentum dependence also shows that hadron-string cascade
does not give strong enough elliptic flows
and underestimates the magnitudes around 1.2-1.5 in minimum bias events,
although it describes the overall trend very well.
Since the elliptic flow is sensitive to the thermaliztion in the early stage,
the above observations indicate
that it is necessary to include additional processes which are effective
in early thermalization than hadron-string cascade processes.

In the analysis of centrality dependence,
we find that the underestimates of $v_2$ in JAM mainly comes from semi-central
collisions ($0.1 < N_{ch}/N_{max} < 0.5$).
It may be possible to interpret this underestimates
as the lack of cooperative processes in JAM.
In semi-central collisions,
number density of produced particles are smaller than in central collisions,
then it is generally more difficult to achive thermalization of the system.
But in the case of cooperative processes such as the phase transition
from superheated liquid to gas,
the transition proceeds catastrophically from a small seed region.

Finally, in the analysis of pseudorapidity dependence,
it is found that the elliptic flows in fragmentation region ($|\eta| > 2$)
are well described in a hadron-string cascade, 
while hydrodynamical description works at mid-pseudorapidities
($|\eta| < 2$).
It suggests that the early thermalization is achieved at mid-rapidities
where there are a huge number of particles produced by mini-jets.

In summary, we find that a hadron-string cascade gives reasonable descriptions 
of elliptic flows up to SPS energies and fragmentation region at RHIC energy,
although it underestimates the elliptic flow at mid-rapidities
especially in semi-central collisions.
The most natural explanation of this underestimates is to assume that
partons produced in mini-jets interact frequently in the early stage
at mid-rapidities and thermalized QGP is formed through these re-scatterings,
which is not included in the present model.
Actually, this is supported by 
the full 3D hydrodynamical model calculation,
which assumes complete local thermalization and reproduces the data
at mid-rapidities.
In order to confirm QGP formation theoretically,
it is desireble to incorporate partonic interactions
in a dynamical model such as JAM.
In this regard, some steps has been taken in the literature~\cite{lin02}, but 
still it is not complete.

\begin{figure}%[htbp]
\centerline{~{\psfig{figure=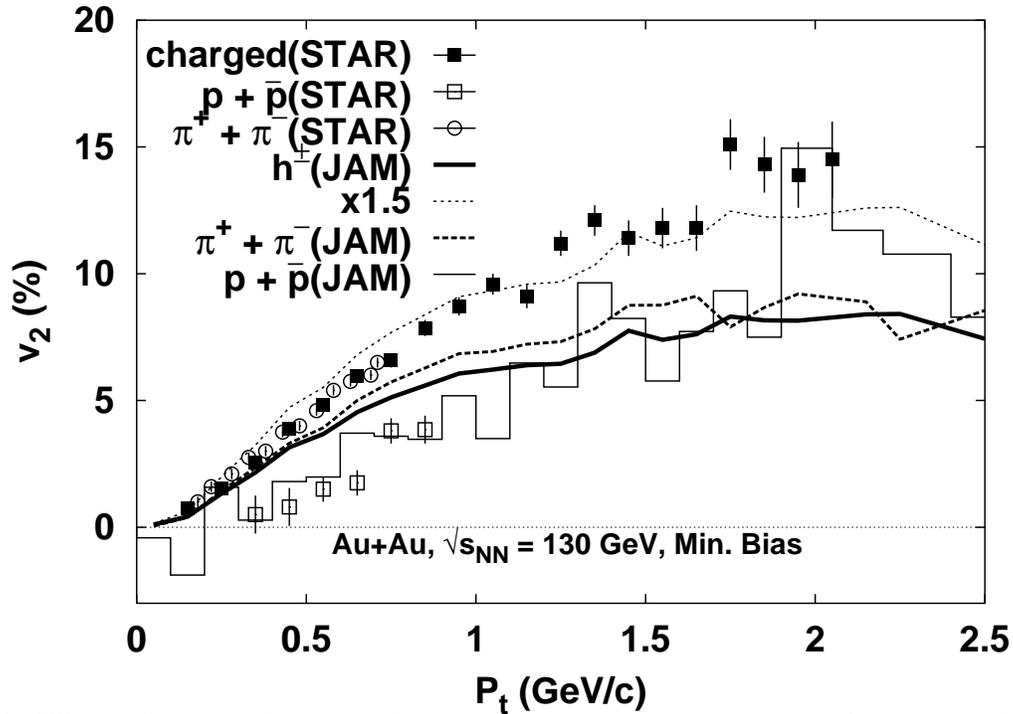,width=14cm,angle=0}}~}
\caption{Elliptic flow as a function of transverse momenta 
for minimum bias events for charged particles}
\label{Fig:Pt}
\end{figure}

\begin{figure}%[htbp]
\centerline{~{\psfig{figure=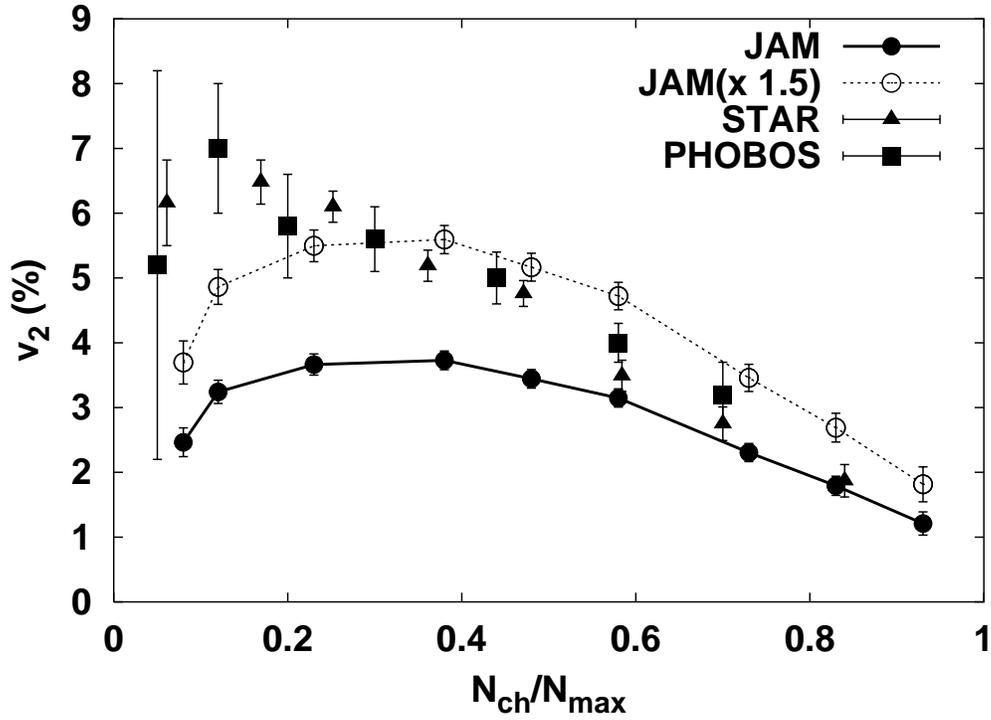,width=14cm,angle=0}}~}
\caption{Elliptic flow as a function of centrality 
for minimum bias events for charged particles}
\label{Fig:Cent}
\end{figure}

%%%%%%%%%%%%%%%%%%%%%%%%%%%%%%%%%%%%
% Possible Modification of Data
%%%%%%%%%%%%%%%%%%%%%%%%%%%%%%%%%%%%
On the other hand, it would be also necessary to make more sophisticated 
analyses in extracting elliptic flows from expeirmental data.
The elliptic flow data were measured using conventional method~\cite{pos98} 
of correlating particles with an event plane.
In this method the observed elliptic coefficient is not correct, if it is
not corrected by dividing by the resolution of the event plane, since 
the observed event plane is not the true reaction plane.
The flow coefficient can be determined without referring to an event plane
by the multiparticle correlation method~\cite{pos01} using cumulants.
The four-particle correlation method is more advantage than two-particle
method, due to elimination of two-particle non-flow effects. 
However in four-particle correlation method, the natural statistical 
errors are larger than the two-particle analysis because of fourth root 
of the result.
We would like to mention here that in Figs.~\ref{Fig:Pt} and \ref{Fig:Cent},
the data may come 
further down due to elimination of non-flow effects, measurements
of elliptic flow by using four-particle correlations method\cite{pos01}. 
In that case, our prediction of 1.5 times the original result of our model
is more encouraging and a positive forward step.

Authors would like to thank Y. Nara for fruitful discussions and Prof. R. K.
Choudhury and Prof. M. Baldo for useful suggestions and comments.

\newpage

\end{document}